\documentclass[12pt]{article}
\begin{document}

\begin{center}
{\Large\bf Gravitational energy in a small region for the modified
Einstein and Landau-Lifshitz pseudotensors}
\end{center}

\begin{center}
Lau Loi So\\
Department of Physics, National Central University, Chung-Li 320, Taiwan\\
(Dated on 23 September 2008)
\end{center}

\begin{abstract}
The purpose of the classical Einstein and Landau-Lifshitz
pseudotensors is for determining the gravitational energy. Neither
of them can guarantee a positive energy in holonomic frames. In
the small sphere approximation, it has been required that the
quasilocal expression for the gravitational energy-momentum
density should be proportional to the Bel-Robinson tensor
$B_{\alpha\beta\mu\nu}$. However, we propose a new tensor
$V_{\alpha\beta\mu\nu}$ which is the sum of certain tensors
$S_{\alpha\beta\mu\nu}$ and $K_{\alpha\beta\mu\nu}$, it has
certain properties so that it gives the same gravitational
``energy-momentum" content as $B_{\alpha\beta\mu\nu}$ does.
Moreover, we show that a modified Einstein pseudotensor turns out
to be one of the Chen-Nester quasilocal expressions, while the
modified Landau-Lifshitz pseudotensor becomes the Papapetrou
pseudotensor; these two modified pseudotensors have positive
gravitational energy in a small region.
\end{abstract}

\section{Introduction}
Gravitational energy should be positive; the proof is not easy. If
the quasilocal gravitational expression is positive on a large
scale that guarantees that it is positive in the small; i.e., the
quasilocal small sphere approximation. Conversely a negative
gravitational energy expression on the small scale implies
non-positive on the large scale. Therefore the small sphere limit
approximation plays a role on testing whether the gravitational
expression has the opportunity to be good (positive energy) or
definitely bad (negative energy).

As there is a successful proof on a large scale \cite{Yau}, we
expect that there should exist at least one small scale
gravitational expression in a holonomic frames.  A good quasilocal
expression should satisfy several requirements: the interior mass
density, the ADM mass \cite{ADM} at the spatial infinity and a
positive small sphere gravitational energy like the Bel-Robinson
tensor $B_{\alpha\beta\mu\nu}$ \cite{Szabados}. This tensor is
desired because it provides a non-negative gravitational
``energy". In vacuum the quasilocal value for the gravitational
energy should be a multiple of
$\frac{4}{3}\pi{}r^{5}B_{\alpha\beta\mu\nu}t^\alpha{}t^\beta{}t^\mu{}t^\nu$
in the small sphere limit approximation, where
$\frac{4}{3}\pi{}r^{3}$ is the Euclidean volume with radius $r$
and $t^\alpha$ is a timelike unit normal \cite{SoCQG}.

The present paper considers two different types of gravitational
quasilocal expressions in holonomic frames.  One is a modified
Einstein pseudotensor which turns out to be one of the Chen-Nester
four quasilocal expressions \cite{Nester}.  The other is a
modified Landau-Lifshitz pseudotensor which is equivalent to the
Papapetrou pseudotensor. Moreover, we propose a new tensor
$V_{\alpha\beta\mu\nu}$ which is a sum of certain tensors,
$S_{\alpha\beta\mu\nu}+K_{\alpha\beta\mu\nu}$, which contributes
the same gravitational ``energy-momentum" value as
$B_{\alpha\beta\mu\nu}$ does.

\section{Ingredients}
Using a Taylor series expansion, the metric tensor can be written
as
\begin{equation}
g_{\alpha\beta}(x)
=g_{\alpha\beta}(0)+(\partial_{\mu}g_{\alpha\beta})(0)x^{\mu}
+\frac{1}{2}(\partial^{2}_{\mu\nu}g_{\alpha\beta})(0)x^{\mu}x^{\nu}+\ldots.
\end{equation}
At the origin in Riemann normal coordinates
\begin{eqnarray}
g_{\alpha\beta}(0)&=&\eta_{\alpha\beta},
\quad\quad\quad\quad\quad\quad\quad\quad~~
\partial_{\mu}g_{\alpha\beta}(0)=0, \\
-3\partial^{2}_{\mu\nu}g_{\alpha\beta}(0)
&=&R_{\alpha\mu\beta\nu}+R_{\alpha\nu\beta\mu},\quad\quad
-3\partial_{\nu}\Gamma^{\mu}{}_{\alpha\beta}(0)
=R^{\mu}{}_{\alpha\beta\nu}+R^{\mu}{}_{\beta\alpha\nu}.
\end{eqnarray}

In vacuum the Bel-Robinson tensor $B_{\alpha\beta\mu\nu}$, and the
tensors $S_{\alpha\beta\mu\nu}$ and $K_{\alpha\beta\mu\nu}$ are
defined as follows
\begin{eqnarray}
B_{\alpha\beta\mu\nu}&:=&R_{\alpha\lambda\mu\sigma}R_{\beta}{}^{\lambda}{}_{\nu}{}^{\sigma}
+R_{\alpha\lambda\nu\sigma}R_{\beta}{}^{\lambda}{}_{\mu}{}^{\sigma}
-\frac{1}{8}g_{\alpha\beta}g_{\mu\nu}R^{2},\\
S_{\alpha\beta\mu\nu}&:=&R_{\alpha\mu\lambda\sigma}R_{\beta\nu}{}{}^{\lambda\sigma}
+R_{\alpha\nu\lambda\sigma}R_{\beta\mu}{}{}^{\lambda\sigma}
+\frac{1}{4}g_{\alpha\beta}g_{\mu\nu}R^{2},\\
K_{\alpha\beta\mu\nu}&:=&R_{\alpha\lambda\beta\sigma}R_{\mu}{}^{\lambda}{}_{\nu}{}^{\sigma}
+R_{\alpha\lambda\beta\sigma}R_{\nu}{}^{\lambda}{}_{\mu}{}^{\sigma}
-\frac{3}{8}g_{\alpha\beta}g_{\mu\nu}R^{2},
\end{eqnarray}
where $R^{2}=R_{\lambda\sigma\rho\tau}R^{\lambda\sigma\rho\tau}$.
In order to extract the vacuum ``energy-momentum" density from the
above three tensors, one can use the analog of the ``electric"
$E_{ab}$ and ``magnetic" $H_{ab}$ parts of the Weyl tensor,
\begin{eqnarray}
E_{ab}=C_{0a0b}, \quad H_{ab}=\ast{C_{0a0b}},
\end{eqnarray}
where $C_{\alpha\beta\mu\nu}$ is the Weyl conformal tensor and
$\ast{C_{\alpha\beta\mu\nu}}$ is its dual,
\begin{equation}
\ast{C_{\alpha\beta\mu\nu}}
=\frac{1}{2}\sqrt{-g}\epsilon_{\alpha\beta\lambda\sigma}
C^{\lambda\sigma}{}_{\mu\nu}.
\end{equation}
In a simple form using the Riemann tensor in vacuum
\begin{equation}
E_{ab}=R_{0a0b},\quad
H_{ab}=\frac{1}{2}R_{0amn}\epsilon_{b}{}^{mn}.
\end{equation}
Certain commonly occurring quadratic combinations of the Riemann
tensor components in terms of the electric $E_{ab}$ and magnetic
$H_{ab}$ parts in vacuum are
\begin{equation}
R_{0a0b}R_{0}{}^{a}{}_{0}{}^{b}=E_{ab}E^{ab},\quad
R_{0abc}R_{0}{}^{abc}=2H_{ab}H^{ab},\quad
R_{abcd}R^{abcd}=4E_{ab}E^{ab}.
\end{equation}
In particular, the Riemann squared tensor can be written in terms
of the electric and magnetic parts as
\begin{equation}
R_{\alpha\beta\mu\nu}R^{\alpha\beta\mu\nu}
=8(E_{ab}E^{ab}-H_{ab}H^{ab}),
\end{equation}
where Greek letters refer to spacetime and Latin stand for space.

The Bel-Robinson tensor in vacuum includes the ``energy-momentum"
\begin{eqnarray}
B_{\mu{}000}=B_{\mu{}0l}{}^{l}=B_{\mu{}l0}{}^{l}
=(E_{ab}E^{ab}+H_{ab}H^{ab},2\epsilon_{c}{}^{ab}E_{ad}H^{d}{}_{b}),
\end{eqnarray}
where $l=1,2,3$.  Likewise for the tensors $S_{\alpha\beta\mu\nu}$
and $K_{\alpha\beta\mu\nu}$:
\begin{eqnarray}
S_{\mu{}000}&=&S_{\mu{}l0}{}^{l}~=~2(E_{ab}E^{ab}-H_{ab}H^{ab},0),\\
S_{\mu{}0l}{}^{l}&=&-10(E_{ab}E^{ab}-H_{ab}H^{ab},0),\\
K_{\mu{}000}&=&K_{\mu{}l0}{}^{l}
~=~(-E_{ab}E^{ab}+3H_{ab}H^{ab},2\epsilon_{c}{}^{ab}E_{ad}H^{d}{}_{b}),\\
K_{\mu{}0l}{}^{l}&=&(11E_{ab}E^{ab}-9H_{ab}H^{ab},2\epsilon_{c}{}^{ab}E_{ad}H^{d}{}_{b}).
\end{eqnarray}
Consequently, there is an identity between the components of
$B_{\alpha\beta\mu\nu}$, $S_{\alpha\beta\mu\nu}$ and
$K_{\alpha\beta\mu\nu}$:
\begin{equation}
B_{\mu{}000}=B_{\mu{}0l}{}^{l}=B_{\mu{}l0}{}^{l}
=S_{\mu{}l0}{}^{l}+K_{\mu{}l0}{}^{l}
=S_{\mu{}0l}{}^{l}+K_{\mu{}0l}{}^{l}
=S_{\mu{}000}+K_{\mu{}000}.\label{20May2008}
\end{equation}
This means that it is not necessary to obtain the Bel-Robinson
tensor $B_{0000}$, $B_{00l}{}^{l}$ or $B_{\mu{}l0}{}^{l}$ for the
positive ``energy" requirement, the sum $S_{0000}+K_{0000}$,
$S_{00l}{}^{l}+K_{00l}{}^{l}$ or
$S_{\mu{}l0}{}^{l}+K_{\mu{}l0}{}^{l}$ can fulfill the same task.

Based on the above argument, we propose a new tensor
$V_{\alpha\beta\mu\nu}$ defined as follows
\begin{eqnarray}
V_{\alpha\beta\mu\nu}&:=&S_{\alpha\beta\mu\nu}+K_{\alpha\beta\mu\nu}\nonumber\\
&=&R_{\alpha\mu\lambda\sigma}R_{\beta\nu}{}{}^{\lambda\sigma}
+R_{\alpha\nu\lambda\sigma}R_{\beta\mu}{}{}^{\lambda\sigma}
+R_{\alpha\lambda\beta\sigma}R_{\mu}{}^{\lambda}{}_{\nu}{}^{\sigma}
+R_{\alpha\lambda\beta\sigma}R_{\nu}{}^{\lambda}{}_{\mu}{}^{\sigma}
-\frac{1}{8}g_{\alpha\beta}g_{\mu\nu}R^{2}.\quad
\end{eqnarray}
This tensor has some nice properties but is not as good as the
Bel-Robinson tensor.  In particular, $B_{\alpha\beta\mu\nu}$ is
completely symmetric but $V_{\alpha\beta\mu\nu}$ is not. Some
detailed properties for $S_{\alpha\mu\nu}$,
$K_{\alpha\beta\mu\nu}$ and $V_{\alpha\beta\mu\nu}$ in vacuum are
\begin{eqnarray}
&&S_{\alpha\beta\mu\nu}\equiv{}S_{(\alpha\beta)(\mu\nu)}\equiv{}S_{(\mu\nu)(\alpha\beta)},\quad~~{}
S_{\alpha\beta\mu}{}^{\mu}\equiv{}\frac{3}{2}g_{\alpha\beta}R^{2},\quad\quad~{}S_{\alpha\mu\beta}{}^{\mu}\equiv{}0,\\
&&K_{\alpha\beta\mu\nu}\equiv{}K_{(\alpha\beta)(\mu\nu)}\equiv{}K_{(\mu\nu)(\alpha\beta)},\quad{}
K_{\alpha\beta\mu}{}^{\mu}\equiv{}-\frac{3}{2}g_{\alpha\beta}R^{2},\quad{}K_{\alpha\mu\beta}{}^{\mu}\equiv{}0,\quad\\
&&V_{\alpha\beta\mu\nu}\equiv{}V_{(\alpha\beta)(\mu\nu)}\equiv{}V_{(\mu\nu)(\alpha\beta)},\quad\quad{}
V_{\alpha\beta\mu}{}^{\mu}\equiv{}0\equiv{}V_{\alpha\mu\beta}{}^{\mu},\quad{}
V_{\mu{}000}\equiv{}V_{\mu{}0l}{}^{l}\equiv{}V_{\mu{}l0}{}^{l}.\quad\quad
\end{eqnarray}

\section{The interior, ADM and gravitational energy}
Three physical regions of interest for the energy of a gravitating
system in general relativity are: the interior mass-energy
density, the ADM mass \cite{ADM} at the spatial infinity, and the
gravitational field energy-momentum in vacuum. Einstein described
gravitational energy by the classical pseudotensor
$t_{\alpha}{}^{\mu}$ which follows from the Freud superpotential
\cite{Freud}
\begin{equation}
U_{\alpha}{}^{[\mu\nu]}=\sqrt{-g}\left(
\delta^{\rho}_{\alpha}\Gamma^{\lambda}{}_{\lambda}{}^{\pi}
+\delta^{\pi}_{\alpha}\Gamma^{\rho\lambda}{}_{\lambda}
+\Gamma^{\pi\rho}{}_{\alpha} \right)\delta^{\mu\nu}_{\rho\pi},
\end{equation}
in a way which guarantees conservation.  Such superpotentials
cannot be uniquely defined, for example suppose
\begin{equation}
t_{\alpha}{}^{\mu}=\partial_{\nu}U_{\alpha}{}^{[\mu\nu]},
\end{equation}
then one can always introduce a new pseudotensor such as
\begin{equation}
\widetilde{t}_{\alpha}{}^{\mu}=t_{\alpha}{}^{\mu}
+\partial_{\nu}\widetilde{U}_{\alpha}{}^{[\mu\nu]},
\end{equation}
because $\partial_{\mu}\widetilde{t}_{\alpha}{}^{\mu}=0$ and
$\partial_{\mu}t_{\alpha}{}^{\mu}=0$ they are both conserved
densities. However, the interior mass density and the ADM mass
energy provide restrictions for some guidance, so that one can
have some more physical energy-momentum components. Consider the
following generalization of the Freud superpotential
\begin{equation}
U_{\alpha}{}^{[\mu\nu]}=\sqrt{-g}\left(
k_{1}\delta^{\rho}_{\alpha}\Gamma^{\lambda}{}_{\lambda}{}^{\pi}
+k_{2}\delta^{\pi}_{\alpha}\Gamma^{\rho\lambda}{}_{\lambda}
+k_{3}\Gamma^{\pi\rho}{}_{\alpha}
\right)\delta^{\mu\nu}_{\rho\pi},
\end{equation}
where $k_{1}$, $k_{2}$ and $k_{3}$ are the extra added constants.
Inside matter at the origin in Riemann normal coordinates to
zeroth order (where $\kappa=8\pi{}G/c^{4}$)
\begin{eqnarray}
2\kappa\,t_{\alpha}{}^{\mu}&=&\frac{1}{3}\sqrt{-g}\left\{
(k_{1}+2k_{2}+3k_{3})R_{\alpha}{}^{\mu}
-(k_{1}+2k_{2})\delta^{\mu}_{\alpha}R\right\} \nonumber\\
&=&2\sqrt{-g}G_{\alpha}{}^{\mu}\nonumber\\
&=&2\kappa\,T_{\alpha}{}^{\mu},
\end{eqnarray}
provided that
\begin{eqnarray}
k_{1}+2k_{2}+3k_{3}&=&6, \label{14aNoc2006}\\
k_{1}+2k_{2}&=&3.        \label{14bNoc2006}
\end{eqnarray}
To check the ADM mass, let's use the Schwarzschild metric in
Cartesian coordinates:
\begin{equation}
ds^{2}=-\left(1-\frac{2GM}{r}\right)dt^{2}
+\left(1+\frac{2GM}{r}\right)(dx^{2}+dy^{2}+dz^{2}).
\end{equation}
The energy-momentum is
\begin{equation}
2\kappa{}P_{\alpha}=-\frac{1}{2}\oint{}U_{\alpha}{}^{[\mu\nu]}\epsilon_{\mu\nu},
\end{equation}
where
$\epsilon_{\mu\nu}=\frac{1}{2}\epsilon_{\mu\nu\lambda\sigma}dx^{\lambda}dx^{\sigma}$.
The associated ADM mass energy term is
\begin{equation}
M=-\oint\frac{1}{4\kappa}U_{0}{}^{[\mu\nu]}\epsilon_{\mu\nu}=\frac{1}{2}\left(k_{1}+k_{3}\right)M,
\end{equation}
which generates one more constraint
\begin{equation}
k_{1}+k_{3}=2.  \label{14cNoc2006}
\end{equation}
Considering (\ref{14aNoc2006}), (\ref{14bNoc2006}) and
(\ref{14cNoc2006}), the unique solution is
\begin{equation}
k_{1}=k_{2}=k_{3}=1.
\end{equation}
Therefore only the Einstein pseudotensor or others which are
asymptotically equivalent, such as the Landau-Lifshitz
pseudotensor, have this property. But the evaluation of the
Einstein and Landau-Lifshitz pseudotensors do not give a positive
gravitational energy \cite{Deser,MTW}. The present paper
introduces the flat metric tensor $\eta_{\alpha\beta}$ along with
$g_{\alpha\beta}$, then it turns out that certain modified
Einstein and Landau-Lifshitz pseudotensors do have positive
gravitational energy in the small region vacuum limit.

\section{Modification of the Einstein pseudotensor}
The modified quasilocal expression in holonomic frames \cite{So2}
is summarized as
\begin{eqnarray}\label{9June2006}
2\kappa\,{\cal{}B}_{c_{1},c_{2}}(N)
&=&2\kappa\,{\cal{}B}_{p}(N)+c_{1}i_{N}\Delta{}\Gamma^{\alpha}{}_{\beta}
\wedge\Delta\eta_{\alpha}{}^{\beta}
-c_{2}\Delta\Gamma^{\alpha}{}_{\beta}
\wedge{}i_{N}\Delta\eta_{\alpha}{}^{\beta} \nonumber\\
&=&-\frac{N^{\alpha}}{2}\left\{_{E}U_{\alpha}{}^{[\mu\nu]}
+c_{1}\sqrt{-g}h^{\lambda\pi}\Gamma^{\sigma}{}_{\alpha\pi}
\delta^{\mu\nu}_{\lambda\sigma}
+c_{2}\sqrt{-g}h^{\beta\sigma}\Gamma^{\tau}{}_{\lambda\beta}
\delta^{\lambda\mu\nu}_{\tau\sigma\alpha}
\right\}\epsilon_{\mu\nu},\quad~~
\end{eqnarray}
where $c_{1},c_{2}$ are real numbers and
$h_{\alpha\beta}:=g_{\alpha\beta}-\eta_{\alpha\beta}$. When
$(c_{1},c_{2})=(0,0),~(0,1),~(1,0)$ and $(1,1)$, this recovers the
original Chen-Nester four holonomic expressions (Note that here we
take $\overline{\Gamma}^{\alpha}{}_{\beta\mu}=0$ so
$\Delta\Gamma^{\alpha}{}_{\beta\mu}=\Gamma^{\alpha}{}_{\beta\mu}$).
The superpotential can be extracted from (\ref{9June2006})
\begin{equation}
U_{\alpha}{}^{[\mu\nu]} ={}_{E}U_{\alpha}{}^{[\mu\nu]}
+c_{1}\sqrt{-g}h^{\lambda\pi}\Gamma^{\sigma}{}_{\alpha\pi}
\delta^{\mu\nu}_{\lambda\sigma}
+c_{2}\sqrt{-g}h^{\beta\sigma}\Gamma^{\tau}{}_{\lambda\beta}
\delta^{\lambda\mu\nu}_{\tau\sigma\alpha},
\end{equation}
where
$_{E}U_{\alpha}{}^{[\mu\nu]}=-\sqrt{-g}g^{\beta\sigma}\Gamma^{\tau}{}_{\lambda\beta}
\delta_{\tau\sigma\alpha}^{\lambda\mu\nu}$ which is the Freud
superpotential.  Note that the $h\Gamma$ terms do not affect the
results inside matter and at spatial infinity, but only the second
order vacuum value. The small region results of the modified
Chen-Nester expressions in compact form in RNC is
\begin{eqnarray}
2\kappa\,t_{\alpha}{}^{\beta}=2G_{\alpha}{}^{\beta}+\frac{1}{18}
\left\{
\begin{array}{cccc}
(4+c_{1}-5c_{2})B_{\alpha}{}^{\beta}{}_{\xi\kappa}\\
-(1-2c_{1}+c_{2})S_{\alpha}{}^{\beta}{}_{\xi\kappa}~~\\
+(c_{1}-3c_{2})K_{\alpha}{}^{\beta}{}_{\xi\kappa}~~~~~~\\
\end{array}
\right\}
x^{\xi}x^{\kappa}+{\cal{}O}({\rm{Ricci}},x)+{\cal{}O}(x^{3}).
\end{eqnarray}
Using a calculation method similar to that in \cite{Nester1}, the
gravitational energy-momentum in the small sphere limit is
\begin{eqnarray}\label{12June2006}
P_{\mu}&=&(-E,\vec{P})\nonumber\\
&=&\frac{1}{2\kappa}\int\frac{1}{18}\left\{
(4+c_{1}-5c_{2})B^{0}{}_{\mu{}ij}
-(1-2c_{1}+c_{2})S^{0}{}_{\mu{}ij}
+(c_{1}-3c_{2})K^{0}{}_{\mu{}ij} \right\}
x^{i}x^{j}d^{3}x \nonumber \\
&=&-\frac{c_{1}}{180G}r^{5}B_{\mu000},
\end{eqnarray}
provided $c_{1}>0$ and we take the unique combination
$c_{1}+2c_{2}=1$, which is the constraint from requiring the
coefficient of $S_{\alpha\beta\mu\nu}$ and $K_{\alpha\beta\mu\nu}$
to be the same. There exists an infinite number of solutions
because of the parameter $c_{1}$. Different solutions are
associated with different boundary conditions. However there is
one solution with locally positive energy and a simple boundary
condition which is when $(c_{1},c_{2})=(1,0)$ in equation
(\ref{9June2006}). It reduces to one of the Chen-Nester holonomic
expressions, ${\cal{}B}_{c}(N)$. In detail
\begin{equation}
2\kappa\,{\cal{}B}_{c}(N)=i_{N}\Gamma^{\alpha}{}_{\beta}\wedge\Delta\eta_{\alpha}{}^{\beta}
+\Delta\Gamma^{\alpha}{}_{\beta}\wedge{}i_{N}\eta_{\alpha}{}^{\beta},
\end{equation}
the associated superpotential is
\begin{equation}\label{9aJune2006}
{\cal{}U}_{\alpha}{}^{[\mu\nu]}={}_{E}U_{\alpha}{}^{[\mu\nu]}
+\sqrt{-g}h^{\lambda\pi}\Gamma^{\sigma}{}_{\alpha\pi}
\delta^{\mu\nu}_{\lambda\sigma}.
\end{equation}
The corresponding gravitational energy-momentum in a small sphere
vacuum region is
\begin{equation}
P_{\mu}=-\frac{1}{180G}r^{5}B_{\mu000}.
\end{equation}
This result shows that the vector $P_{\mu}$ is future pointing and
non-spacelike.

\section{Modification of the Landau-Lifshitz pseudotensor}
The superpotential for the Papapertrou pseudotensor \cite{P
superpotential} is
\begin{eqnarray}
_{P}H^{[\mu\nu][\alpha\beta]}
=\sqrt{-g}(\eta^{\mu\alpha}g^{\nu\beta}-\eta^{\nu\alpha}g^{\mu\beta}
+\eta^{\nu\beta}g^{\mu\alpha}-\eta^{\mu\beta}g^{\nu\alpha}).
\end{eqnarray}
In another form it is
\begin{eqnarray}
_{P}U^{\alpha[\mu\nu]}&=&\partial_{\beta}\left(_{P}H^{[\mu\nu][\alpha\beta]}\right)\nonumber\\
&=&{}_{B}U^{\alpha[\mu\nu]}-\sqrt{-g}\left(
g^{\lambda\sigma}h^{\pi\beta}\Gamma^{\alpha}{}_{\lambda\pi}\delta^{\nu\mu}_{\sigma\beta}
+g^{\alpha\beta}h^{\pi\sigma}\Gamma^{\tau}{}_{\lambda\pi}\delta^{\lambda\mu\nu}_{\tau\sigma\beta}
\right),
\end{eqnarray}
where
$_{L}U^{\alpha[\mu\nu]}=\sqrt{-g}_{B}U^{\alpha[\mu\nu]}=gg^{\alpha\beta}
g^{\pi\sigma}\Gamma^{\tau}{}_{\lambda\pi}
\delta_{\tau\sigma\beta}^{\lambda\mu\nu}$, ``L" stands for
Landau-Lifshitz and ``B" refers to Bergmann-Thomson. Once again,
the $h\Gamma$ terms do not affect the results inside matter and at
spatial infinity, they would contribute, however, to the second
order vacuum value.  The pseudotensor can be obtained as
\begin{equation}
2\kappa\,t^{\alpha\beta}=\partial_{\mu}U^{\alpha[\beta\mu]},
\end{equation}
the expression in RNC is
\begin{equation}
2\kappa\,t^{\alpha\beta}=2G^{\alpha\beta}+\frac{1}{9}
\left(4B^{\alpha\beta}{}_{\xi\kappa}
-S^{\alpha\beta}{}_{\xi\kappa}
-K^{\alpha\beta}{}_{\xi\kappa}\right)
x^{\xi}x^{\kappa}+{\cal{}O}({\rm{Ricci}},x)+{\cal{}O}(x^{3}).
\end{equation}
The associated gravitational energy-momentum in the small sphere
vacuum approximation is
\begin{eqnarray}
P_{\mu}&=&\frac{1}{2\kappa}\int\frac{1}{9}\left(
4B^{0}{}_{\mu{}ij}-S^{0}{}_{\mu{}ij}-K^{0}{}_{\mu{}ij}
\right)x^{i}x^{j}d^{3}x \nonumber\\
&=&-\frac{1}{180G}r^{5}B_{\mu{}000}.
\end{eqnarray}
As mentioned previously, the vector $P_{\mu}$ is future pointing
and non-spacelike.

\section{Conclusion}
In the small sphere approximation, the quasilocal expression for
the gravitational energy-momentum density should be proportional
to the Bel-Robinson tensor $B_{\alpha\beta\mu\nu}$. However, we
propose a new tensor $V_{\alpha\beta\mu\nu}$ which is the sum of
the tensors $S_{\alpha\beta\mu\nu}+K_{\alpha\mu\nu}$ which gives
the same gravitational ``energy-momentum" content as
$B_{\alpha\beta\mu\nu}$ does.

The purpose of the classical Einstein and Landau-Lifshitz
pseudotensors are to define the gravitational energy.  Neither of
them can guarantee positive energy in holonomic frames in the
small sphere limit, i.e., $B_{0000}$ in vacuum to second order. In
order to achieve the desired expression for the gravitational
energy, one may consider the modification of these two
pseudotensors. We found that the modified Einstein pseudotensor
becomes one of the Chen-Nester quasilocal expressions, the
modified Landau-Lifshitz pseudotensor is equivalent to the
Papapetrou pseudotensor; these two modified pseudotensors have a
positive gravitational energy which comes from $V_{0000}$ in the
small region vacuum limit.

\section*{Acknowledgment}
This work was supported by NSC 94-2112-M-008-038.


\begin{thebibliography}{3}

\bibitem{Yau}
R. Schoen and S.T. Yau, {\it{}Phy. Rev. Letters} {\bf{}43}, 1457
(1979)

\bibitem{ADM}
R. Arnowitt, S. Deser and C.W. Misner, {\it{}Phys. Rev.}
{\bf{}122}, 997 (1961)

\bibitem{Szabados}
L.B. Szabados, Living Rev. Relativity, {\bf 7} 4 (2004)\\
http://www.livingreviews.org/lrr-2004-4

\bibitem{SoCQG}
L.L. So, {\it Class. Quantum Grav.} {\bf{}25}, 175012 (2008)

\bibitem{Nester}
C.M. Chen and J.M. Nester, {\it{} Class. Quantum Grav.} {\bf{}16},
1279 (1999).

\bibitem{Freud}
Ph. Freud, {\it{}Ann. Math}, {\bf 40}, 417 (1939)

\bibitem{Deser} S. Deser, J.S. Franklin and D. Seminaea,
{\it Class. Quantum Grav.} {\bf 16}, 2815 (1999)

\bibitem{MTW}
C.W. Misner, K.S. Thorne and J.A. Wheeler 1973 {\it{}Gravitation}
(San Francisco, CA: Freeman)

\bibitem{So2}
L.L. So, {\it{}Int. J. Mod. Phys. D} {\bf{}16}, 875 (2007)

\bibitem{Nester1}
L.L. So and J.M. Nester, 2006 Preprint gr/qc0612061

\bibitem{P superpotential}
A. Papapetrou, {\it Einstein's theory of gravitation and flat
space}, Proc. Roy. Irish. Acad.  {\bf A52}, 11 (1948)

\end{thebibliography}
\end{document}